\documentclass[aps,showpacs,preprintnumbers,amsmath,amssymb,nofootinbib]{revtex4}

\usepackage{graphicx}
\usepackage{color}

\def \be  {\begin{equation}}
\def \ee  {\end{equation}}
\def \ee  {\end{equation}}
\def \bea {\begin{eqnarray}}
\def \eea {\end{eqnarray}}

\def\be {\begin{equation}}
\def\ee {\end{equation}}
\def\bea {\begin{eqnarray}}
\def\eea {\end{eqnarray}}
\def\bc {\begin{center}}
\def\ec {\end{center}}
\def\bfg {\begin{figure}}
\def\efg {\end{figure}}
\def\bi {\begin{itemize}}
\def\ei {\end{itemize}}

\def\la {\label}
\def\le {\left}
\def\ri {\right}

\def\a  {\alpha}

\def\e  {\epsilon}

\def\beq{\begin{equation}}
\def\eeq{\end{equation}}
\def\br{\begin{eqnarray}}
\def\er{\end{eqnarray}}
\newcommand{\eel}[1] {\label{#1}\end{equation}}

\begin{document}

\preprint{ECTP-2012-03}

\title{Effects of the Generalized Uncertainty Principle on Compact Stars}

\author{Ahmed Farag Ali}
\email{ahmed.ali@fsc.bu.edu.eg}
\email{ahmed.ali@uleth.ca}
\affiliation{Department of Physics, Faculty of Science, Benha University, Benha 13518, Egypt}

\author{Abdel Nasser Tawfik}
\email{a.tawfik@eng.mti.edu.eg}
\email{atawfik@cern.ch}
\affiliation{Egyptian Center for Theoretical Physics (ECTP), MTI University, Cairo, Egypt}
\affiliation{Research Center for Einstein Physics, Freie-University Berlin, Berlin, Germany}

\date{\today}

\begin{abstract}

Based on the generalized uncertainty principle (GUP), proposed by some approaches to quantum gravity such as string theory and doubly special relativity theories, we investigate the effect of GUP on the thermodynamic properties of compact stars with two different components. We note that the existence of quantum gravity correction tends to resist the collapse of stars if
the GUP parameter $\a$ is taking values between Planck scale and electroweak scale. Comparing with approaches, it is found that the radii of compact stars are found smaller. Increasing energy almost exponentially decreases the radii of compact stars.

\end{abstract}


\maketitle


\section{Introduction}
\label{sec:intr}

Different approaches for the quantum gravity have been proposed to provide a set of predictions for a minimum measurable length, a maximum observable momentum and an essential modification of the Heisenberg uncertainty principle (GUP) \cite{guppapers, BHGUP,kmm,kempf, brau, Scardigli}. The latter is based on modification in the fundamental commutation relation $[x_i, p_j]$.  According to string theory (ST) and black hole physics (BHP), the modification is  proportional to a quadratic moment \cite{guppapers, BHGUP,kmm,kempf, brau, Scardigli}. Based on doubly spacial relativity (DSR), a proportionality to first order momenta (linear) has been suggested \cite{cg}. Different minimal length scale scenarios inspired by various approaches to the quantum gravity have been reviewed in
\cite{Hossenfelder:2012jw}. The approach of Ali et al. \cite{advplb,Ali:2010yn, Das:2010zf}, which shall be utilized in this work, predicts maximum observable momenta and minimal measurable length. It is consistent with DSR, ST and BHP. Furthermore, it ensures that $[x_i,x_j]=0=[p_i,p_j]$ through Jacobi identity. Accordingly, a modification of the uncertainty principle near the Planck energy scale reads
\bea
[x_i,p_j] &=& i \hbar \left[\delta_{i j} - \alpha \left(p
\delta_{i j} + \frac{p_i p_j}{p}\right) +  \alpha^2 \left(p^2
\delta_{i j} + 3 p_i p_j\right) \right]. \label{comm01}
\eea
Apparently, this suggests a modification of the physical momentum \cite{advplb,
Ali:2010yn, Das:2010zf}
\bea
p_i = p_{0i} \le( 1 - \alpha
p_0+2\alpha^2 p_0^2 \ri)~, \label{mom1}
\eea
while $x_i = x_{0i}$ with $x_{0i}, p_{0j}$ satisfying the canonical commutation
relations $ [x_{0i}, p_{0j}] = i \hbar~\delta_{ij}, $ such that
\hbox{$p_{0i}=-i\hbar \partial/\partial{x_{0i}}$}, where $p_{0}$ is
the momentum at low energy satisfying Eq. (\ref{comm01}).

The parameter $\alpha=\alpha_0/(c\, M_{pl}) =\alpha_0 l_{pl}/\hbar$, where $c$, $\alpha_0$, $M_{pl}\, (l_{pl})$ are speed of light, dimensionless parameter of order one, and Planck mass (length), respectively.

In terms of the gravitational constant $G$, the Planck mass and length are given as $M_{pl}=\sqrt{\hbar c /G}$
and $l_{pl}=\sqrt{G \hbar/c^3}$, respectively. Accordingly, the
dispersion relation is modified to the first order of $\alpha$ as follows.
\bea
E^2 = p_0^2\, c^2 \left(1-2\,\alpha\,p_0\right) + M^2\, c^4. \la{modis}
\eea

Together with the modified energy-momentum conservation used in \cite{LV}, this equation introduces a deformation of local Lorentz invariance.  This means that the bounds predicted from the effective field theory framework \cite{Maccione:2007yc} will not apply in our considered model.

The upper bounds on the parameter $\alpha$ has been derived in \cite{Ali:2011fa}
and it was found that it could predict an intermediate length scale between
Planck scale and electroweak scale.
It was suggested that these bounds can be measured using quantum optics techniques
in \cite{Nature} which is considered as milestone in the quantum gravity phenomenology.

Because of the modified physical momentum of Eq.(\ref{mom1}), the classical Hamiltonian as
well as the quantum Hamiltonian should be modified  which affects a lot of quantum phenomena,
and we think it is important to make a quantitative study of these
effects which would open an interesting window for quantum gravity phenomenology especially if
the GUP parameter $\a$ would take values between Planck scale and electroweak scale.
In a series of papers, the effects of GUP on atomic, condensed matter
systems, quark gluon plasma, preheating phase of the universe, inflationary era of the universe and
black holes production at LHC have been  investigated
\cite{dvprl,Ali:2010yn,dvcjp, Ali:2011fa, Elmashad:2012mq, Chemissany:2011nq, Tawfik:2012he, Ali:2012mt}.

In this paper, we continue our investigations to study the impact of quantum gravity effects on the ground state properties of a Fermi gas composed of $N$ ultra-relativistic electrons. It was found by Chandrasekhar \cite{Chandrasekhar:1931ih} that the white dwarfs can be described by ultra-relativistic ideal Fermi gas at which quantum gravity effects could be considerable. Therefore, we study the effect of GUP on the number density, energy density and pressure. We discuss  two kinds of the compact stars. The first one is a white dwarf, in which  the mass contribution is mainly from the cold nuclei. The second kind is where the mass is mainly constituted of Fermi gas. In context of compact stars, the earliest proposal on subatomic stars dates to 1970, when It was proposed that there might exist compact stellar objects that might entirely made of very massive quarks, rather than confined baryons \cite{itoh70}.


\section{GUP Effect on ground state of ultra-relativistic Fermi gas}
\label{sec:mdl}

For an isolated macroscopic body consisting of $N$ non-interacting  and ultra-relativistic particles,
the background of the particles motion is assumed to be flat. We study the
ground state properties of a Fermi gas composed of $N$ ultra-relativistic
electrons, for which the state energy $\epsilon$ is entirely given
by $c p$ i.e., temperature is much larger than the particle's rest mass.
However, at vanishing or very  low temperatures, the vacuum effect of
fermions can be neglected i.e., the total particle number is conserved.
On the other hand, based on GUP  the particles move in quantized gravitational
background. The  modified state density due to GUP has been derived using Liouville theorem \cite{faragali}.

\begin{figure}[htb!]
\includegraphics[width=8.cm,angle=0]{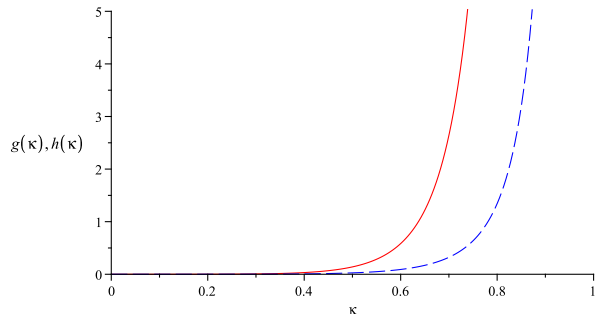}
\caption{Left pane: the dimensionless quantities $g$ (solid curve) and $h$ (dashed curve) are presented as functions of $\kappa$.
}
\label{afig1}
\end{figure}

At vanishing temperature, The modified number of particle of Fermi gas can be given as
\bea \label{eq:pFermi1}
N(p) &=& \frac{8 \pi}{h^3}V \int_0^{p_F} \frac{p^2 dp }{(1-\alpha p)^4},
\eea
where $p_F$ gives the Fermi momentum. Therefore, Eq. (\ref{eq:pFermi1})
can be re-written in terms of  Fermi energy $\epsilon_F$
\bea
N(\epsilon_F) &=& \frac{8 \pi}{3 (h c)^3}\, V\, \frac{\epsilon_F^3}{\left(1-\frac{\alpha}{c} \epsilon_F\right)^3}.
\eea
Introducing $\kappa= \epsilon_F/\epsilon_H$, which is equivalent to $(\alpha/c) \epsilon_F$, then
\bea \label{eq:nkappa1}
N(\kappa) &=& \frac{8 \pi}{(h c)^3}\, V\, \epsilon_H^3\, f(\kappa),
\eea
where $\epsilon_H=c/\alpha$ is the Hagedorn energy and
\bea \label{eq:fkappa1}
f(\kappa) &=& \frac{1}{3} \frac{\kappa^3}{(1-\kappa)^3}.
\eea
Now, we can calculate the average distance between particles
\bea
\bar{d} = \left(\frac{V}{N(\kappa)}\right)^{1/3} &=& \frac{h c}{(8 \pi)^{1/3}} \frac{1}{\epsilon_H} f(\kappa)^{-1/3} =
\pi^{\frac{2}{3}}  \Delta_{min} f(\kappa)^{-1/3}.
\eea
where $\Delta_{min}= \a_0\, l_{pl}$. Then defining,
\bea \label{eq:deltaaa}
\delta &\equiv &\frac{\Delta_{min}}{\bar{d}} = \pi^{-\frac{2}{3}} f(\kappa)^{1/3},
\eea
the ground state energy can be calculated from
\bea \label{eq:tU}
U_0(\epsilon) &=& \frac{8 \pi }{(h c)^3} V \int_0^{\epsilon_F} \frac{\epsilon^3 d\epsilon}{\left(1-\frac{\alpha}{c} \epsilon\right)^4}.
\eea
In terms of $\kappa$, the ground state energy and pressure read
\bea
U_0(\kappa) 
&=& \frac{8 \pi}{(h c)^3}\,  V\, \epsilon_H^4\, g(\kappa), \label{eq:U1}\\
P(\kappa) &=& \frac{N}{V} \e_F - \frac{U_0}{V} =\frac{8 \pi}{(h c)^3} \, \epsilon_H^4\, h(k), \label{eq:P1}
\eea
where
\bea
g(\kappa)&=&\ln(1-\kappa) + \frac{\kappa}{(1-\kappa)^3} - \frac{15}{6} \frac{\kappa^2}{(1-\kappa)^3} + \frac{11}{6} \frac{\kappa^3}{(1-\kappa)^3}, \label{gk} \\ h(\kappa)&=& \frac{1}{3} \frac{\kappa^4}{(1-\kappa)^3} -  \left[\ln(1-\kappa) + \frac{\kappa}{(1-\kappa)^3} - \frac{15}{6} \frac{\kappa^2}{(1-\kappa)^3} + \frac{11}{6} \frac{\kappa^3}{(1-\kappa)^3}\right].
\label{hk}
\eea
These two quantities are presented in Fig. \ref{afig1}. It is obvious that both of them diverge at $\kappa\rightarrow 1$. $g(\kappa)$ diverges much faster than $h(\kappa)$. This would mean that the validity of this approach is limited to the Fermi energy. It is bounded from above by a maximum energy bound $(c/ \alpha)$. This is completely consistent with the predicted maximum measurable momentum ($1/\alpha$) in \cite{advplb,
Ali:2010yn, Das:2010zf}.

When the Fermi energy turns to be much smaller than $\epsilon_H=c/\alpha$, i.e., $\kappa\ll 1$, the distribution function given in Eq. (\ref{eq:fkappa1}) cab be approximated as
\bea \label{eq:fkappa2}
f(\kappa) &=&  \frac{\kappa^3}{3}+ \kappa^4.
\eea
Furthermore, from Eq. (\ref{eq:fkappa2}), it is clear that $\delta \ll 1$.
In this limit, the state number reads
\bea \label{eq:nkappa2}
N(\kappa) &= & \frac{8 \pi}{(h c)^3}\, V\, \epsilon_H^3\, \left(\frac{1}{3} \kappa^3 + \kappa^4\right).
\eea
In this limit, $\kappa$ can be given in terms of number density $N/V$,
\bea \label{eq:kappaDE}
\kappa^3\left(\frac{1}{3} + \kappa\right) &=& \frac{(h c)^3}{8 \pi}\, \frac{1}{\epsilon_H^3}\, \frac{N}{V},
\eea
which would lead to a relation between $\kappa$ and $\delta$
\be
\kappa= (3 \pi^2)^{\frac{1}{3}}\, \delta\, \le[1 - (3 \pi^2)^{\frac{1}{3}}\, \delta\ri] + O(\delta^3) \la{kapaadelta}
\ee

The Hagedorn energy (or equivalently temperature) is defined as \hbox{$\epsilon_H = M_p c^2/\alpha_0$}. It is a scale to set the limit of  applying the GUP approach. Accordingly, Eqs. (\ref{eq:U1}) and (\ref{eq:P1}) can be re-written  as
\bea
U_0(\kappa) &=& \frac{8 \pi V}{(h c)^3} \epsilon_H^4 \left[\frac{\kappa^4}{4}+ \frac{4 \kappa^5}{5}\right], \label{eq:U2}\\
P(\kappa) &=& \frac{8 \pi}{(h c)^3} \epsilon_H^4 \le[ \frac{\kappa^4}{12}+\frac{\kappa^5}{5}\ri], \label{eq:P2}
\eea
where  the condition $\kappa\ll 1$ has been  assumed.
The corrections to $U_0(\kappa)$ and $P(\kappa)$ would be in terms of $\delta$
\bea
\frac{U_0(\kappa)}{V} &=& \frac{3^{4/3}}{4} \frac{h c}{(8 \pi)^{\frac{1}{3}}} \le(\frac{N}{V}\ri)^{4/3} \le(1+
\frac{16}{5} (3 \pi^2)^{1/3}\delta\ri), \label{eq:U3}\\
P(\kappa)
 &=&  \frac{3^{1/3}}{4} \frac{h c}{(8 \pi)^{\frac{1}{3}}} \le(\frac{N}{V}\ri)^{4/3} \le(1+ \frac{12}{5} (3 \pi^2)^{1/3}\delta\ri), \label{eq:P3}
\eea
where Eq. (\ref{kapaadelta}) has been utilized.

We should note that the results are for a framework that has modified dispersion relations of Eq. (\ref{modis}) following from the GUP and a modified measure on momentum space of Eq. (\ref{eq:pFermi1}). It is therefore outside the effective
field theory framework \cite{Maccione:2007yc}.
%
%

\section{Compact Stars at vanishing temperature}
\label{sec:cs}

\subsection{Non-relativistic cold nuclei with $M=2 N/m_p$}

It is conjectured that the major contribution to the mass of
white dwarfs, for instant, is coming from the non-relativistic cold
nuclei having mass $M=2 N/m_p$, where  we consider a star with
mass M and N electrons, and  $m_p$ is the mass of the proton \cite{Chandrasekhar:1931ih}.
The white dwarfs have two properties \cite{wd};
the first one is that the electrons are described by relativistic dynamics and
the second one is that the electron gas is completely degenerate.
Based on these properties, the electron gas would be treated as a zero temperature gas.
In this case $\epsilon_F=2  N c^2/\alpha_0 m_p$
indicating that $\kappa\ll 1$ and Eq. (\ref{eq:P2}) seems to reflect
that the quantum gravity effects increase the degenerate pressure.
Would this effects is confirmed, then the quantum gravity corrections to the mass of white
dwarfs arise.

In order to keep the electron gas at a given density,
the electronic degeneracy pressure is supposed to resist
the gravitational collapse. At equilibrium, the pressure reads
\bea \label{eq:pequil}
P_0(R) &=& \frac{\lambda}{4 \pi} G \left(\frac{M}{R^2}\right)^2,
\eea
where $G$ is the gravitational constant, $R^3\equiv V$ and $\lambda$
is free parameter of the order of unity. Nevertheless, its value depends on how the matter is distributed inside the white dwarf. From Eqs. (\ref{eq:P3}) and (\ref{eq:pequil}), and by ignoring the constants of unity, the pressure can be expressed in terms of the internal energy
\bea
\le(\frac{N}{V}\ri)^{4/3} \le( 1 + \delta\ri) = \frac{G M^2}{R^4}.
\eea
By substituting $M=2 N m_p$, the correction to the mass of the compact star would be
\bea
M = M_0 \le(1+ \left(\frac{N}{V}\right)^{\frac{1}{3}} \a \hbar\ri),
\label{result}
\eea
where we used the relation given in Eq. (\ref{eq:deltaaa}) in order to obtain $\delta= \le(N/V\ri)^{\frac{1}{3}} \a \hbar$. Furthermore, we set \hbox{$M_0=\le(h c/G\ri)^{\frac{3}{2}} (2 m_p)^{-2}$}.
%

Eq. (\ref{result}) presents an interesting result. The quantum gravity correction seems to be  proportional to the density number of the star.  Let us consider a physical example for a white dwarf, in which the density number $N= 10^{36}$, the average distance $\bar{d}= 10^{-12}$, and the Fermi energy $\e_F= 10^{5}$ eV.

In various derivations of the GUP (see e.g. refs.\cite{guppapers,BHGUP}), $\alpha_0$ is normally assumed to be $O(1)$.This can be thought of as the lower bound (the physics becomes even more ill understood for sub-Planckian, $\alpha_0<1$ regimes).  Our analysis uses an upper bound, and hence a length scale $\alpha_0\times$(Planck length) which is intermediate between the Planck and the electroweak scale (which it cannot exceed on phenomenological grounds). We used the bound $\alpha\leq 10^{-2}$ GeV$^{-1}$ (i.e. $\alpha_0 \leq 10^{17}$) depending on the derived bounds on the parameter $\alpha_0$ in \cite{Ali:2011fa}. This bound was derived by calculating the effect of quantum gravity with non relativistic heavy meson systems like charmonium \cite{Ali:2011fa} which may be a relevant example for the white dwarfs which is constituted mainly from non-relativistic nuclei.


Based on these values, the quantum gravity correction to the mass of the white dwarf is given by

\bea
M_{GUP}= M_0\le(1 + 10^{-5}\ri).
\eea

Two remarks are in order now. The first one is that the correction seems to be more stringent than the one derived in previous work about compact stars with quantum gravity corrections \cite{wang}. The correction given in Ref. \cite{wang} is $10^{-10}$ . The second remark is that the quantum gravity corrections in our work is positive referring to resisting the collapse of the compact stars. It is obvious that this conclusion agrees with the result in \cite{wang}.

\subsection{Ultra-relativistic nuclei with $M=U_0/c^2$}

\begin{figure}[htb!]
\includegraphics[width=8.cm,angle=0]{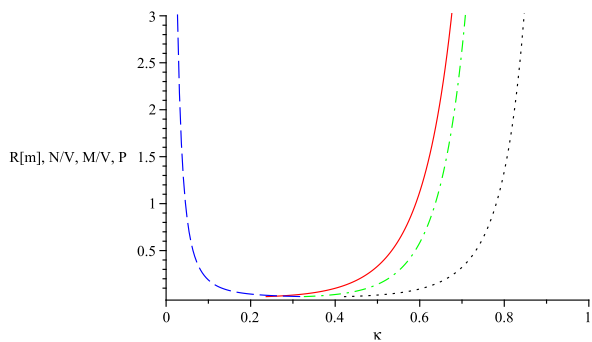}
\caption{The modified radius of white dwarf, Eq. (\ref{eq:Rr}), is given in dependence on $\kappa$ (dashed curve) at $\alpha_0 \approx 10^{17}$ i.e $\alpha\approx10^{-2}~$GeV$^{-1}$ \cite{Ali:2011fa}. The modified  normalized particle density in Fermi gas at vanishing temperature $N(\kappa)/V$  is given in dependence on $\kappa$ (solid curve) at $\alpha=10^{-2}~$GeV$^{-1}$. The normalized mass density $M(\kappa)/V$ is given as dash-dotted curve at $\alpha=10^{-2}~$GeV$^{-1}$. The normalized pressure is given as dotted curve.
}
\label{afig4}
\end{figure}

There are other configurations in which the star is almost composed of ultra-relativistic
nuclei. In this case, the mass of the nuclei is compressed as $M=U_0/c^2$.
The constituents of the white dwarfs are characterized by an ideal Fermi gas and total mass $M=U_0/c^2$.  The electronic degeneracy pressure is assumed to resist the gravitational collapse, Eq. (\ref{eq:pequil}). At equilibrium, the radius of the white dwarf  is given by
\bea
\label{eq:Rpropto1}
R &=& 
\frac{\lambda}{8 \pi} R_S Q(\kappa), \label{eq:Rr}
\eea
where $Q(\kappa)=g(\kappa)/h(\kappa)$ and the parameter $\lambda$ approximately equals unity.  In the considered case, the Schwarzschild radius reads
\bea
R_S &=& 2 \,G\, \frac{M}{c^2} = 2 \,G\, \frac{U_0}{c^4}.
\eea
At $\lambda\approx1$, the results are presented in Fig. \ref{afig4} for the stringent value of the parameter $\alpha$ \cite{Ali:2011fa}. We observe that the radius approaches its minima as  $\kappa\rightarrow 1$ , and it divers as $\kappa\rightarrow 0$. The number density $N/V$ from Eq. (\ref{eq:nkappa1})  and mass density $M= U_0/c^2$ from  Eq. (\ref{eq:U1}) are presented also in Fig. \ref{afig4}. We observe that the number density, the mass density and the pressure
approach their minima as $\kappa\rightarrow 0$, but they reach their maximum values as $\kappa\rightarrow 1$.

Current observations indicate that white dwarfs have smaller radii than expected \cite{Mathews:2006nq}. The behavior of $R$ vs $\kappa$ in Fig. (\ref{afig4})  suggests that $R$ is decreasing as $\kappa\rightarrow 1$ , lead to a possible explanation for the smaller radii observations. Similar analysis has been done in the context of doubly special relativity and modified dispersion relations \cite{Camacho:2006qg, Gregg:2008jb}. The calculations of the neutron star mass in DSR were done in \cite{AmelinoCamelia:2009tv}.

\section{conclusions}

We investigate the effect of GUP on the thermodynamical properties
of the compact stars and study the impact on the Chandrasekhar
limit and gravitational collapse. We found that the quantum gravity
corrections would increase the Chandrasekhar limit and hence they resist
the gravitational collapse. Besides, we found that the radius of
the compact star is decreasing as the energy increasing which might be
considered as a possible explanation for the smaller radii observations.

\section*{Acknowledgments}
The research of AFA is supported by Benha University. The research of AT has been partly supported by the German--Egyptian Scientific Projects (GESP ID: 1378). AFA and AT like to thank Prof. Antonino Zichichi for his kind invitation to attend the International School of Subnuclear Physics 2012 at the ''Ettore Majorana Foundation and Centre for Scientific Culture'' in Erice-Italy, where the present work was started. The authors gratefully thank the anonymous referee
for useful comments and suggestions which helped to improve the paper.



\end{document}